\pdfoutput=1
\RequirePackage{ifpdf}
\ifpdf 
\documentclass[pdftex]{sigma}
\else
\documentclass{sigma}
\fi

\numberwithin{equation}{section}

\numberwithin{theorem}{section}
\numberwithin{proposition}{section}

\begin{document}

\newcommand{\arXivNumber}{1412.2867}

\allowdisplaybreaks

\renewcommand{\thefootnote}{$\star$}

\renewcommand{\PaperNumber}{045}

\FirstPageHeading

\ShortArticleName{Non-Integrability of Some Higher-Order Painlev\'e
Equations in the Sense of Liouville}

\ArticleName{Non-Integrability of Some Higher-Order\\ Painlev\'e
Equations in the Sense of Liouville\footnote{This paper is a~contribution to the Special Issue on
Algebraic Methods in Dynamical Systems.
The full collection is available at
\href{http://www.emis.de/journals/SIGMA/AMDS2014.html}{http://www.emis.de/journals/SIGMA/AMDS2014.html}}}

\Author{Ognyan CHRISTOV~$^\dag$ and Georgi GEORGIEV~$^\ddag$}

\AuthorNameForHeading{O.~Christov and G.~Georgiev}

\Address{$^\dag$~Faculty of Mathematics and Informatics, Sofia University, 1164 Sofia, Bulgaria}
\EmailD{\href{mailto:christov@fmi.uni-sofia.bg}{christov@fmi.uni-sofia.bg}}

\Address{$^\ddag$~Department of Mathematics and Informatics,
University of Transport, 1574, Sofia, Bulgaria}
\EmailD{\href{mailto:ggeorgiev6@yahoo.com}{ggeorgiev6@yahoo.com}}

\ArticleDates{Received December 10, 2014, in f\/inal form June 10, 2015; Published online June 17, 2015}

\Abstract{In this paper we study the equation
$$
w^{(4)} = 5 w'' (w^2 - w') + 5 w (w')^2 - w^5 + (\lambda z +
\alpha)w + \gamma,
$$
which is one of the higher-order Painlev\'e equations (i.e.,
equations in the polynomial class having the Painlev\'e property).
Like the classical Painlev\'e equations, this equation admits a
Hamiltonian formulation, B\"{a}cklund transformations and families
of rational and special functions.
We prove that this equation considered as a Hamiltonian system
with parameters $\gamma/\lambda = 3 k$, $\gamma/\lambda = 3 k - 1$, $k
\in \mathbb{Z}$, is not integrable in Liouville sense by means of
rational f\/irst integrals. To do that we use the Ziglin--Morales-Ruiz--Ramis approach.
Then we study the integrability of the second and  third members
of the $\mathrm{P}_{\mathrm{II}}$-hierarchy. Again as in the
previous case it turns out that the normal variational equations
are particular cases of the generalized conf\/luent hypergeometric
equations whose dif\/ferential Galois groups are non-commutative and
hence, they are obstructions to integrability. }


\Classification{70H05; 70H07; 34M55; 37J30}

\renewcommand{\thefootnote}{\arabic{footnote}}
\setcounter{footnote}{0}

\section{Introduction}

The Painlev\'e property for a system of dif\/ferential equations is
the property that its general solution is without movable critical
points.

Let us given a nonlinear dif\/ferential equation
\begin{gather*}
F \big(x, y, y', \ldots, y^{(n)}\big) = 0, \qquad x, y \in \mathbb{CP}^1,
\end{gather*}
where $F$ is a polynomial with respect to $y, y', \ldots,
y^{(n)}$. Let $y = \varphi (x)$ be a solution of the above
equation, which usually turns out to be a multivalued holomorphic
function.

A {\it critical point} is a point of ramif\/ication of the solution
$y = \varphi (x)$. The critical point is called {\it movable} if
its position depends on the solution $\varphi$, that is, on the
constants of integration. For example, the f\/irst order equations
without movable critical points are the linear equations and the
Riccati equations (see \cite{Gol,Zo}). Equations with
this property are called equations of Painlev\'e type.

In the beginning of 20th century Painlev\'e and Gambier
investigated this property for second-order ordinary dif\/ferential
equations. They proved that there are f\/ifty equations possessing
the Painlev\'e property. Among them six equations turned out to be
new at that time, now called classical Painlev\'e equations.
Although derived in a pure mathematical way, the six Painlev\'e
equations have appeared in many physical applications: in the
description of nonlinear waves, in statistical mechanics, in the
theory of quantum gravity, in topological f\/ield theory, in plasma
physics, in the theory of random matrix models and so on.

The classical Painlev\'e equations have many remarkable
properties, in particular they admit a Hamiltonian formulation. In
\cite{MR2} Morales-Ruiz asked the question about the non-integrability
as Hamiltonian systems of classical Painlev\'e equations which
have particular rational solutions. This question was answered
af\/f\/irmatively for $\mathrm{P}_{\mathrm{II}}$ with values of the
parameter $\alpha = n \in \mathbb{Z} $. For the recent development
in the study of the non-integrability of the other Painlev\'e
equations we refer to~\cite{TSto}. In a very recent
paper~\cite{ZF} \.{Z}o{\l}\c{a}dek and Filipuk have proved that the
classical Painlev\'e equations do not admit a f\/irst integral that
can be expressed in terms of elementary functions, except for some
known cases of $\mathrm{P}_{\mathrm{III}}$ and
$\mathrm{P}_{\mathrm{V}}$.

It is natural to extend the question of non-integrability to the
higher-order Painlev\'e equations. Consider the following
fourth-order nonlinear ordinary dif\/ferential equation
\begin{gather}
\label{1.1}
w^{(4)} = 5 w'' \big(w^2 - w'\big) + 5 w (w')^2 - w^5 +
(\lambda z + \alpha)w + \gamma,
\end{gather}
where $\lambda$, $\alpha$, $\gamma$ are complex parameters.

This equation appears as a group-invariant reduction of the
modif\/ied  Kaup--Ku\-per\-shmidt (or Sawada--Kotera) equation, see
for instance \cite{Hone,Kudr1}. Then it appears as equation F-XVIII in the
classif\/ication made by  Cosgrove~\cite{Cos} of all fourth- and
f\/ifth-order equations with Painlev\'e property. It is also studied
by Gromak~\cite{Gro} from dif\/ferent points of view. It is
proven in~\cite{Kudr1}, that (\ref{1.1}) with $\lambda = 1$, $\alpha
= 0$ has no polynomial f\/irst integral.

Like the classical Painlev\'e equations, this equation admits a
Hamiltonian formulation, B\"{a}ck\-lund transformations and families
of rational and special functions. For instance:
\begin{itemize}\itemsep=0pt

\item
when $\lambda = 0$, $\gamma \neq 0 $ it is solved in terms of
hyperelliptic functions;

\item
when  $\lambda = 0$, $\gamma = 0 $ it is solved via elliptic
functions;

\item
when  $\gamma = -\lambda/2$, $w (z)$ can be expressed in terms of
two Painlev\'e I solutions.
\end{itemize}

Further, we assume that $\lambda \neq 0$.

The equation (\ref{1.1}) possesses two families of rational
solutions:

I) $\gamma/\lambda = 3 k$, $k \in \mathbb{Z}$
\begin{gather*}
k = 0, \quad w = 0; \qquad k = 1, \quad w_{(1)} = - \frac{3
\lambda}{\alpha + \lambda z} ; \qquad \text{etc}.
\end{gather*}

II)  $\gamma/\lambda = 3 k - 1$, $k \in \mathbb{Z}$
\begin{gather*}
k = 0, \quad  w = \frac{\lambda}{\alpha + \lambda z};
\qquad  k = 1, \quad w_{(1)} = - \frac{2 \lambda}{\alpha + \lambda z}
; \qquad \text{etc}.
\end{gather*}
In fact, these two families are the only rational solutions of~(\ref{1.1})~\cite[Theorem~3]{Gro}.

Denote $q_1 (z) := w (z)$, $\varepsilon^2 = 1$. Then the equation
(\ref{1.1}) can be presented as two equivalent $2 + 1/2$ degrees
of freedom Hamiltonian systems with
\begin{gather*}
H_{\varepsilon} = \frac{1}{2} p_2 ^2 +
\frac{7-9\varepsilon}{12}q_2 ^3 + p_1 q_2 -
\frac{1+3\varepsilon}{4} p_1 q_1 ^2 + \frac{3\varepsilon-1}{4} q_2
(\lambda z + \alpha) + \left(\gamma + \frac{3\varepsilon-1}{4}
\lambda \right) q_1 .
\end{gather*}
The corresponding system of equations is ($'= d/dz$):
\begin{alignat}{3}
& q_1 '   =   q_2 - \frac{3\varepsilon+1}{4} q_1 ^2, \qquad  && p_1 '  =  \frac{1+3\varepsilon}{2} p_1 q_1 - \gamma - \frac{3\varepsilon -1}{4} \lambda,  & \nonumber\\
& q_2 '   =   p_2, \qquad &&  p_2 '  =  -p_1 -
\frac{7-9\varepsilon}{4} q_2 ^2 - \frac{3\varepsilon-1}{4}
(\lambda z + \alpha).& \label{1.5}
\end{alignat}

There exist B\"{a}cklund transformations (see \cite{Gro}) $T_1$, $T_2$ and $T := T_2 T_1$ for the equa\-tion~(\ref{1.1}) acting on the parameters in the following way:
\begin{gather}
T_1 (\lambda) = \lambda, \qquad T_1 (\alpha) = \alpha,
\qquad T_1 (\gamma) = - \gamma - \lambda,\nonumber
\\
\label{1.7} T_2 (\lambda) = \lambda, \qquad T_2 (\alpha) = \alpha,
\qquad T_2 (\gamma) =  -\gamma + 2 \lambda,\nonumber
\\
\label{1.71}
 T (\gamma) = \gamma + 3 \lambda, \qquad T^{-1} (\gamma) = \gamma - 3 \lambda.
\end{gather}
Gromak has shown that these B\"{a}cklund transformations are
birational. It is easy to see that they are canonical also.

We can extend in a natural way the Hamiltonian system (\ref{1.5})
to a three degrees of freedom autonomous system by denoting
$\hat{H} (q_1, q_2, z, p_1, p_2, F) := H_{\varepsilon} + F $. Then
we have
\begin{alignat}{3}
& \frac{d q_1}{d s}   =   q_2 - \frac{3\varepsilon+1}{4}
q_1 ^2, \qquad &&
\frac{d p_1}{d s} = \frac{1+3\varepsilon}{2} p_1 q_1 - \gamma - \frac{3\varepsilon -1}{4} \lambda, & \nonumber\\
& \frac{d q_2}{d s}   =   p_2, \qquad &&
\frac{d p_2}{d s}  =  -p_1 - \frac{7-9\varepsilon}{4} q_2 ^2 - \frac{3\varepsilon-1}{4} (\lambda z + \alpha), & \label{1.8} \\
& \frac{d z}{d s}     =   1, \qquad &&  \frac{d F}{d
s} = - \lambda \frac{3\varepsilon-1}{4} q_2 . & \nonumber
\end{alignat}

Our f\/irst result is the following
\begin{theorem}\label{th1}
The Hamiltonian system \eqref{1.8} with parameters
$\gamma/\lambda = 3 k$, $\gamma/\lambda = 3 k - 1$, $k \in \mathbb{Z}$
is not integrable in the Liouville sense by means of rational
first integrals.
\end{theorem}

We say that a Hamiltonian system with $n$ degrees of freedom is
integrable in the sense of Liouville if there exist~$n$ (almost
everywhere) independent f\/irst integrals in involution.

We use as a tool the Ziglin--Morales-Ruiz--Ramis theorem. We obtain
a particular solution of~(\ref{1.8}), write the normal variational
equation and study its dif\/ferential Galois group which appears to
be large.

The paper is organized as follows. In Section~\ref{section2} we summarize the
necessary facts about the integrability of Hamiltonian systems and
the theory of linear equations with singular points and its
relation to the dif\/ferential Galois theory. In \cite{Katz} N.~Katz
and O.~Gabber have calculated the Galois groups of some classes of
linear equations using purely algebraic arguments~-- global
characterization of semisimple algebras. In Section~\ref{section3} we apply
their result to prove Theorem~\ref{th1}. In Section~\ref{section4} we recover
 Katz's result about the  Galois group for our particular linear equation by giving its topological generators.
It turns out that the linear equations which appear here are the
conf\/luent generalized hypergeometric equations. We use the
approach of Duval and Mitschi~\cite{DM} for the calculation of the
formal monodromy, exponential torus and Stokes matrices in the
corresponding case.

In fact, the conf\/luent generalized hypergeometric equations have
appeared also along the study of other higher-order Painlev\'e
equations.  In Section~\ref{section5} we prove that the second and the third
members of the $\mathrm{P}_{\mathrm{II}}$-hierarchy are
non-integrable in the Hamiltonian context for some particular
values of the parameters. Again the dif\/ferential Galois groups of
conf\/luent ge\-ne\-ra\-li\-zed hypergeometric equations are obstructions to
integrability. We conjecture that the higher members in the
$\mathrm{P}_{\mathrm{II}}$-hierarchy satisfy the same property.

\section{Theory}\label{section2}

In this section we recall some notions and facts about
integrability of Hamiltonian systems in the complex domain, the
Ziglin--Morales-Ruiz--Ramis theory and their relations with
dif\/ferential Galois groups of linear equations.

 Consider a Hamiltonian system
\begin{gather}
\label{2.1}
\dot{x} = X_{H} (x), \qquad t \in \mathbb{C}, \qquad x
\in M
\end{gather}
corresponding to an analytic Hamiltonian $H$, def\/ined on the
complex $2 n$-dimensional mani\-fold~$M$. Suppose the system
(\ref{2.1}) has a non-equilibrium solution~$\Psi (t)$. Denote by
$\Gamma$ its phase curve. We can write the equation in variation
(VE) along this solution
\begin{gather}
\label{2.2}
\dot{\mathbf{\xi}} = D X_{H} ( \Psi (t)) \mathbf{\xi},
\qquad \mathbf{\xi} \in T_{\Gamma} M.
\end{gather}

Further, consider the normal bundle of $\Gamma$, $F:= T_{\Gamma} M
/ TM$ and let $\pi \colon T_{\Gamma} M \to F$ be the natural
projection. The equation~(\ref{2.2}) induces an equation on~$F$
\begin{gather}
\label{2.3} \dot{\eta} = \pi_{*} (D X_{H} ( \Psi (t))\big(\pi^{-1}
\eta\big) , \qquad \eta \in F,
\end{gather}
which is  called the normal variational equation (NVE) along
$\Gamma$. The NVE (\ref{2.3}) admits a~f\/irst integral~$d H$,
linear on the f\/ibers of $F$. The level set $F_p := \{\eta \in F \,|\,
d H (\eta) = p \}$, $p \in \mathbb{C}$,
 is $(2 n - 2)$-dimensional af\/f\/ine bundle over~$\Gamma$.
We shall call $F_p$ the reduced phase space of~(\ref{2.3}) and the
restriction of the NVE on~$F_p$ is called the reduced normal
variational equation.

Then the main result of the Ziglin--Morales-Ruiz--Ramis \cite{MR1}
theory is:
\begin{theorem}\label{th2}
Suppose that the Hamiltonian system \eqref{2.1} has
$n$ meromorphic first integrals in involution. Then the identity
component~$G^0$ of the Galois group of the variational equation is
abelian.
\end{theorem}

Next we consider a linear non-autonomous system
\begin{gather}
\label{2.10}
y' = A (x) y, \qquad y \in \mathbb{C}^n ,
\end{gather}
or equivalently a linear homogeneous dif\/ferential equation
\begin{gather}
\label{2.11}
y^{(n)} + a_1 (x) y^{(n-1)} + \cdots + a_n (x) y = 0,
\end{gather}
with $x \in \mathbb{CP}^1$ (which is enough for our purposes) and
$A \in \mathrm{gl} (n, \mathbb{C} (x))$, $a_j (x) \in \mathbb{C}
(x)$. Let $S:=\{x_1, \ldots, x_s\}$ be the set of singular points
of~(\ref{2.10}) (or~(\ref{2.11})) and let $Y (x)$ be a fundamental
solution of~(\ref{2.10})  (or~(\ref{2.11}))  at $x_0 \in
\mathbb{C} \setminus S$. By the existence theorem  there is a~fundamental solution~$Y (x)$, analytic in a vicinity of~$x_0$. The
continuation of~$Y (x)$
 along a nontrivial loop on $\mathbb{CP}^1$ def\/ines  a linear automorphism
of the vector space of all solutions analytic in the neighborhood
of~$x_0$, called the monodromy transformation. Analytically this
transformation can be presented in the following way. The linear
automorphism $\Delta_{\gamma}$, associated with a loop $\gamma \in
\pi_1 (\mathbb{CP}^1 \setminus S, x_0)$ corresponds to
multiplication of~$Y (x)$ from the right by a constant matrix
$M_{\gamma}$, called monodromy matrix
\begin{gather*}
\Delta_{\gamma} Y (x) = Y (x) M_{\gamma}.
\end{gather*}
The set of these matrices form the monodromy group. Equivalently,
the monodromy group can be def\/ined as a group of automorphisms of
the solution space~\cite{vPS,Zo}.

We may attach another object to the system (\ref{2.10}) (or
(\ref{2.11}))~-- a dif\/ferential Galois group. A~dif\/ferential f\/ield~$K$ is a f\/ield with a derivation~$\partial = {}'$, i.e., an~additive
mapping satisfying Leibnitz rule. A dif\/ferential automorphism of
$K$ is an auto\-mor\-phism commuting with the derivation.

The coef\/f\/icient f\/ield in (\ref{2.10}) (and~(\ref{2.11})) is $K =
\mathbb{C} (x)$. Let $y_{i j}$ be the elements of the fundamental
matrix $Y (x)$. Let $L (y_{i j})$ be the extension of $K$
generated by $K$ and $y_{i j}$~-- a~dif\/ferential f\/ield. This
extension is called a Picard--Vessiot extension.
 Similarly to classical Galois theory we def\/ine the Galois group
$G := \operatorname{Gal}_{K} (L) = \operatorname{Gal} (L/K)$ to be the group of all dif\/ferential
auto\-morphisms of $L$ leaving the elements of $K$ f\/ixed. The Galois
group is, in fact, an algebraic group. It has a unique connected
component~$G^0$ which contains the identity and which is a~normal
subgroup of f\/inite index. The Galois group $G$ can be represented
as an algebraic linear subgroup of $\mathrm{GL} (n, \mathbb{C})$
by
\begin{gather*}
\sigma (Y (x)) = Y (x) R_{\sigma},
\end{gather*}
where $\sigma \in G$ and $R_{\sigma} \in \mathrm{GL} (n,
\mathbb{C})$.

We can do the same locally at $a \in \mathbb{CP}^1$, replacing
$\mathbb{C} (x)$ by the f\/ield of germs of meromorphic functions at
$a$. In this way we can speak of a local dif\/ferential Galois group
$G_a$ of (\ref{2.10}) at $a \in \mathbb{CP}^1$, def\/ined in the
same way for Picard--Vessiot extensions of the f\/ield $\mathbb{C}
\{x-a\}[(x-a)^{-1}]$.

One should note that by its def\/inition the monodromy group is
contained in the dif\/ferential Galois group of the corresponding
system.

We say that two linear systems $y' = A (x) y$ and $z' = B (x) z$
are $K$-equivalent if the latter is obtained from the f\/irst by a
$K$-linear change $y = P z$, $P \in \mathrm{GL} (n, \mathbb{C})$ and
$B = P^{-1} A P - P^{-1} P' $.

Next, we review some facts from the theory of linear systems with
singularities. We call a~singular point~$x_i$ {\it regular} if any
of the solutions of~(\ref{2.10}) (or of~(\ref{2.11})) has at most
polynomial growth in arbitrary sector with a vertex at~$x_i$.
Otherwise the singular point is called {\it irregular}.

We say that the system (\ref{2.10}) has a singularity of the Fuchs
type at $x_i$ if $A (x)$ has a simple pole at $x = x_i$. For the
equation~(\ref{2.11}) the Fuchs type singularity at  $x_i$ means
that the functions $(x - x_i)^j a_j (x)$ are holomorphic in a
neighborhood of~$x_i$.

If the system (\ref{2.10}) has a singularity of the Fuchs type,
then  this singularity is regular. The converse is not true.
However, for the equation~(\ref{2.11}) the regular singularities
coincide with the singularities of the Fuchs type.

\begin{theorem}[\cite{Schl}]\label{th3}
For a system with only
regular singular points, the differential Galois group coincides
with the Zariski closure in $\mathrm{GL}(n,\mathbb{C})$ of the
mo\-no\-dro\-my group.
\end{theorem}

Now we brief\/ly recall the Ramis description of the local Galois
group of (\ref{2.10}) at 0 which we assume to be an irregular
singularity. To the end of the section we follow mainly Mitschi
(see \cite[pp.~368--370]{Mi2} and  \cite[pp.~153--159]{Mi3}).

Let $K = \mathbb{C}\{x\}[x^{-1}] (
\widehat{K}=\mathbb{C}[[x]][x^{-1}])$ be the f\/ield of convergent
Laurent series near $0$ (f\/ield of formal Laurent series), $K_t =
\mathbb{C}\{t\}[t^{-1}] ( \widehat{K}=\mathbb{C}[[t]][t^{-1}])$
are the same objects with respect to the variable $t$ and $A \in
\mathrm{gl} (n, K)$. It is known from the classical  theory that
there exists a formal fundamental solution to~(\ref{2.11}):
\begin{gather}
\label{2.12}
\widehat{Y} (t) = \widehat{H} (t) x^L e^{Q (t)} ,
\end{gather}
where $t^{\sigma} = x$, $\sigma \in \mathbb{N}^*$, $L = \operatorname{Mat} (n,
\mathbb{C})$, $\widehat{H} \in \mathrm{GL} (n, \widehat{K}_t)$ and
$Q = \mathrm{diag} (q_1, \ldots, q_n)$, $q_i \in t^{-1}
\mathbb{C}[\frac{1}{t}]$, $i = 1, \ldots, n$. The integer $\sigma$
is called ramif\/ication degree at~$0$. Denote also $\zeta = e^{2
\pi i/\sigma}$.

First we recall the formal invariants of~(\ref{2.11}). The change
of variable $x \to x e^{2 \pi i}$ commutes with the derivation, so
it def\/ines an element $\widehat{m} \in G$, the formal monodromy
($t \to t \zeta$ commutes with the corresponding derivation).
Relative to $\widehat{Y}$, the automorphism $\widehat{m}$ can be
represented by a matrix~$\widehat{M}$:
\begin{gather*}
\widehat{Y} (t \zeta) = \widehat{Y} (t) \widehat{M} .
\end{gather*}

By def\/inition the exponential torus $\mathcal{T}$ of (\ref{2.10})
relative to $\widehat{Y}$ is the group of the dif\/ferential
$\widehat{K}_t$-automorphisms of the dif\/ferential extension{\samepage
\begin{gather*}
\widehat{K}_t \big(e^Q\big) = \widehat{K}_t \big(e^{q_1}, e^{q_2}, \ldots,
e^{q_n}\big) \qquad \mathrm{of} \quad \widehat{K}_t.
\end{gather*}
$\mathcal{T}$ is isomorphic to $(\mathbb{C}^*)^l$, where $l$ is
the rank of $\mathbb{Z}$-module generated by the $q_i$'s.}

The matrix $\widehat{M}$, clearly invariant by
$\widehat{K}$-equivalence is a formal invariant of (\ref{2.11}).
The same thing applies to the exponential torus $\mathcal{T}$.

Let $V_d (\alpha)$ be an open sector in $\mathbb{C}^* \setminus
\{0\}$ with its vertex at $0$:
\begin{gather*}
V_d (\alpha) = \left\{ x \in \mathbb{C}^* \,|\, 0 < |x| < R, \, d -
\frac{\alpha}{2} < {\mathrm{arg}} (x) <  d + \frac{\alpha}{2} \right\}
\end{gather*}
and let $f$ be a holomorphic function on $V_d (\alpha)$. We say
that $f$ is asymptotic to $\hat{f} = \sum\limits_{n=0} ^{\infty} a_n x^n
\in \mathbb{C} [[x]]$ on $V_d (\alpha)$ (in Poincar\'e sense) if,
for every closed subsector $W \subset V_d (\alpha)$ there exists a~positive constant $M_{W, n}$, such that for every $x \in W$
\begin{gather*}
| x |^{-n} \left|f (x) - \sum_{m=0} ^{n-1} a_m x^m \right| \leq M_{W,
n}
\end{gather*}
for every $n$. We write $f \sim \hat{f}$ on~$V_d (\alpha)$.

Let us restrict ourselves to the case when all non-zero
expressions $(q_i - q_j)$ have the same degree, that is, $(q_i -
q_j) = (\lambda_i - \lambda_j) t^{-k}$, $i, j = 1, \ldots, n$, $k \in \mathbb{N}^*$. By the classical theory   \cite{MRa1, Sib,Wasow} we have the
following result:
\begin{theorem}\label{th4}
For the system \eqref{2.10} with a formal solution
\eqref{2.12}, there exists an actual solution $Y = H x^L e^Q$,
where $H \in \mathrm{GL} (n, \mathbb{C}\{t\})$ has asymptotic
expansion $\widehat{H}$ $(H \sim \widehat{H}$ and $Y \sim
\widehat{Y})$ in any open angular sector with opening $\pi/(k
\sigma)$.
\end{theorem}

We want to extend the solution $Y$ to sectors with opening greater
than $\pi/(k \sigma)$. For this purpose we def\/ine:
\begin{itemize}\itemsep=0pt
\item a Stokes ray as a direction where, for some $i, j = 1, \ldots, n$, one has
$ \mathrm{Re}[q_i (t) - q_j (t)] = 0$.

\item a singular ray is  a direction of maximal decay for some $\exp(q_i - q_j)$,
i.e., a bisecting ray of a maximal sector where $\mathrm{Re}[q_i
(t) - q_j (t)] < 0$.
\end{itemize}

Let $d$ be a singular direction for (\ref{2.10}) at $x = 0$, let
$d^+$ and $d^-$ be nearby directions with arguments $d^+ = d +
\varepsilon, d^- = d - \varepsilon$. Then $V^{\pm} = V_{d^{\pm}}
(\pi/(k\sigma) )$ are two overlaping sectors contai\-ning~$d$. Let
$Y^{-}$ and $Y^{+}$ be actual solutions of~(\ref{2.10}), such that
$Y^{-} \sim \widehat{Y}$ in~$V^{-}$ and~$Y^{+} \sim \widehat{Y}$
in~$V^{+}$. Hence, we have two actual solutions $Y^{-}, Y^{+}$
over $V_d (\pi/(k \sigma))$ (by analytic continuation to this
sector, $\varepsilon \to 0$). Then there exists $S_d \in
\mathrm{GL} (n, \mathbb{C})$, such that
\begin{gather*}
Y^{-} = Y^{+} S_d.
\end{gather*}
The matrix $S_d$ is called Stokes matrix (or multiplier) with
respect to $d$ and $\widehat{Y}$. The Stokes matrices are
unipotent. Moreover, they are invariant under $K$-equivalence,
that is, they are analytic invariants for~(\ref{2.10}) (see also~\cite{BJL}).

The actual solutions $Y$ are usually obtained by summation
procedure of~$\widehat{H}$ along non-singular directions in
maximal sectors. We will not recall here the summation theory
developed by Ramis (see, for instance~\cite{MRa1,vPS} for more
details) because in our particular case there exist fundamental
systems of solutions near the irregular point in suitable sectors
expressed by Meijer $G$-function~\cite{Meijer}.

Finally, we have a theorem that generalize the Schlesinger's
result for the Fuchsian case.
\begin{theorem}[Ramis]\label{th5}
With respect to the formal solution
\eqref{2.12} the analytic Galois group of~\eqref{2.11} at~$0$ is
the  Zariski closure in~$\mathrm{GL}(n,\mathbb{C})$ of the
subgroup
 generated by the formal monodromy~$\widehat{M}$, the exponential torus~$\mathcal{T}$ and
 the Stokes matrices~$S_d$ for all singular rays.
\end{theorem}

Now, let $G_a$ be the local Galois groups of (\ref{2.10}), $a \in
S$. All~$G_a$ can be simultaneously identif\/ied with closed
subgroups of $G$ and the following result holds
\cite[Proposition~1.3]{Mi2}:

{\it  The global Galois group $G$ is topologically generated in $\mathrm{GL} (n, \mathbb{C})$
 by the subgroups~$G_a$, for all $a \in S$.}

\section{Proof of Theorem~\ref{th1}}\label{section3}

Consider f\/irst the family of rational solutions (I). Take $\gamma/\lambda = 0$ or $k = 0$ and
$w=0$. Then it is straightforward to be seen that
\begin{gather}
w = q_1 = 0, \qquad  q_2 = 0, \qquad p_2 = 0, \qquad p_1 = \frac{1-3\varepsilon}{4} (\lambda s + \alpha), \nonumber\\
 z = s, \qquad F = F_0 = \operatorname{const},
\label{3.1}
\end{gather}
is a particular solution.

Let $\xi_j = d q_j$, $\eta_j = d p_j$, $j = 1, 2$, $\xi_3 = d s$, $\eta_3 = dF$ be the variations. The variational equation
along the solution (\ref{3.1}) takes the form
\begin{alignat*}{3}
& \xi_1 '  =  \xi_2,  \qquad && \eta_1 '  =   \frac{1+3\varepsilon}{2} p_1 \xi_1,&  \nonumber \\
& \xi_2 '   =   \eta_2, \qquad && \eta_2 '  =  - \eta_1 + \frac{1-3\varepsilon}{4} \lambda \xi_3 , &  \nonumber\\
& \xi_3 '   =   0, \qquad &&  \eta_3 '  =  - \lambda \frac{3 \varepsilon - 1}{4} \xi_2 .& 
\end{alignat*}
Then the normal variational equation   is
\begin{gather}
\xi_1 '   =   \xi_2, \qquad \eta_1 '  =   \frac{1+3\varepsilon}{2} p_1 \xi_1 , \qquad
\xi_2 '   =   \eta_2, \qquad \eta_2 ' =  - \eta_1 .\label{3.3}
\end{gather}
Reducing (\ref{3.3}) to a single equation yields
\begin{gather*}
\xi_1 ^{(4)} = (\lambda z + \alpha) \xi_1 .
\end{gather*}
After setting $z = 1/\tau$ we obtain
\begin{gather*}
y ^{(4)} + \frac{12}{\tau} y ^{(3)} + \frac{36}{\tau^2} y ^{(2)} + \frac{24}{\tau^3} y ' - \frac{\lambda + \alpha \tau}{\tau^9} y = 0
\end{gather*}
from where $\tau = 0$ or $z = \infty$
 is an irregular singular point. After changing the independent variable
$z \to \lambda z + \alpha$ we get ($\partial = d/dz$)
\begin{gather}
\label{3.5}
L_1 \xi_1 = 0, \qquad L_1 = \partial ^4 + a z, \qquad a: = - 1/\lambda^4 .
\end{gather}
The operator $L_1$ is usually called an Airy type operator. It is irreducible and it is obviously invariant
under $\partial \to - \partial$, hence, $L_1$ is selfdual in the terminology of~\cite{Katz}.
In the same paper Katz \cite{Katz} has found that the identity component  of the Galois group of $L_1 \xi_1 = 0$
is $G^0 = \mathrm{Sp} (4, \mathbb{C})$ which is clearly non-commutative.
 Therefore, by  Theorem \ref{th2} the Hamiltonian
system~(\ref{1.8}) is not integrable in a neighborhood of the particular solution~(\ref{3.1}).

Further we consider the second family of rational solutions (II). Take $\gamma/\lambda = -1$ or  $k = 0$ and $w = \frac{1}{z + \alpha/\lambda}$.
Then we have the following particular solution to (\ref{1.8})
\begin{gather}
w = q_1   =   \frac{1}{s + \frac{\alpha}{\lambda}}, \qquad  q_2 = \frac{3}{4} \frac{\varepsilon-1}{(s + \frac{\alpha}{\lambda})^2}, \qquad
 p_2 = -\frac{3}{2} \frac{\varepsilon-1}{(s + \frac{\alpha}{\lambda})^3}, \nonumber \\
 p_1   =    \frac{1-3\varepsilon}{4} (\lambda s + \alpha), \qquad  z = s, \qquad F = \frac{3 \lambda}{4} \frac{1-\varepsilon}{s + \frac{\alpha}{\lambda}} +  F_0 .\label{3.6}
\end{gather}
The VE along this solution is
\begin{alignat*}{3}
& \xi_1 '   =   \xi_2  - \frac{1+3\varepsilon}{2} q_1 \xi_1, \qquad &&
\eta_1 '  =   \frac{1+3\varepsilon}{2} (p_1 \xi_1 + q_1 \eta_1), & \nonumber \\
& \xi_2 '   =   \eta_2, \qquad &&
\eta_2 '  =  - \eta_1 + \frac{1-3\varepsilon}{4} \lambda \xi_3 + \frac{9\varepsilon-7}{2} q_2 \xi_2, & \\
& \xi_3 '   =   0, \qquad &&
\eta_3 '  =  - \lambda \frac{3 \varepsilon - 1}{4} \xi_2 .&
\end{alignat*}
Then the NVE becomes
\begin{alignat}{3}
& \xi_1 '   =   \xi_2 - \frac{1+3\varepsilon}{2} q_1 \xi_1 , \qquad && \eta_1 '  = \frac{1+3\varepsilon}{2} (p_1 \xi_1 + q_1 \eta_1) , & \nonumber \\
& \xi_2 '   =   \eta_2, \qquad && \eta_2 ' =  - \eta_1 + \frac{9\varepsilon-7}{2} q_2 \xi_2 .& \label{3.8}
\end{alignat}
Again reducing (\ref{3.8}) to a single equation gives
\begin{gather*}
\xi_1 ^{(4)} - \frac{10}{(z + \frac{\alpha}{\lambda})^2} \, \xi_1 '' + \frac{20}{(z + \frac{\alpha}{\lambda})^3}   \xi_1 '
- \left[ \frac{20}{(z + \frac{\alpha}{\lambda})^4} + \lambda \left( z + \frac{\alpha}{\lambda} \right) \right]   \xi_1 = 0.
\end{gather*}
We make some transformations in order to put this equation in appropriate form. First we shift the independent variable
$z + \frac{\alpha}{\lambda} \to z$. Then we make a change
$x = \frac{\lambda z^5}{5^4}$, which is a f\/inite branched covering map $\mathbb{CP}^1 \to \mathbb{CP}^1$.
In general, the dif\/ferential Galois group is changed under such transformation, but the identity component
remains unchanged (see \cite[p.~28]{MR1}).
As a~result we get
\begin{gather*}
\xi_1 ^{(4)} + \frac{24}{5 x}   \xi_1 ^{(3)} + \frac{86}{25 x^2}   \xi_1 '' + \frac{4}{125 x^3}   \xi_1 '
- \left(\frac{4}{125 x^4} + \frac{1}{x^3}\right)   \xi_1 = 0.
\end{gather*}
Finally, we put $u := x^{2/5} \xi_1$ from where denoting $\delta = x d/d x$ we obtain
\begin{gather}
\label{3.10}
L_2 u := \delta \left(\delta - \frac{2}{5} -1\right) \left(\delta + \frac{1}{5} -1\right) \left(\delta + \frac{2}{5} - 1\right) u - x u = 0.
\end{gather}
The operator $L_2$ in (\ref{3.10}) is a particular case of so-called Kloosterman operators
\begin{gather*}
\mathrm{Kl} = \prod_1 ^n ( \delta - a_i) + \lambda x
\end{gather*}
with $n = 4$, $\lambda = -1$, $a_1 = 0$, $a_2 = 7/5$, $a_3 = 4/5$, $a_4 =
3/5$. Katz has found that (see \cite[Theorem~4.5.3, pp.~59--60]{Katz}) the identity component of the Galois group of $L_2 u = 0$ is  $G^0 =  \mathrm{Sp} (4, \mathbb{C})$
which is noncommutative.
Hence, by Theorem \ref{th2} the Hamiltonian system~(\ref{1.8})
is not integrable in a neighborhood of the particular solution~(\ref{3.6}).

To f\/inish the proof note that  having the variable $q_1 = w$ we can obtain the
other phase variables from the equations~(\ref{1.5}).
Recall that the equation~(\ref{1.1}) has rational solutions only for
$\gamma/\lambda = 3 k$ and $\gamma/\lambda = 3 k-1$, $k \in \mathbb{Z}$ and  therefore,
the Hamiltonian system~(\ref{1.8}) has particular rational solutions for these values of the parameters.

For the f\/irst family (I) $\gamma/\lambda = 3 k$, $k \in \mathbb{Z}$
we can relate the solution $w=0$ for $\gamma/\lambda = 0 (k=0)$ and the corresponding rational solution~$w_{(k)}$ for
$\gamma/\lambda = 3 k$ via the B\"{a}cklund transformation~$T^k$, $k \in \mathbb{Z}$~(\ref{1.71}).
Since these transformations acting on the phase coordinates are birational (and canonical), the
non-integrability of the Hamiltonian system~(\ref{1.8}) for $\gamma/\lambda = 0$ $(k=0)$ by means of rational f\/irst integrals
implies the non-integrability of the corresponding Hamiltonian systems for $\gamma/\lambda = 3 k$, $k$ is any integer.
Applying the same arguments to the rational solutions of the second family (II),
 we conclude the non-integrability  of the Hamiltonian systems
for $\gamma/\lambda = 3 k-1$, $k \in \mathbb{Z}$.
This ends the proof of Theorem~\ref{th1}.

\section{Stokes matrices}\label{section4}

In this section we explicitly compute the dif\/ferential Galois group of (\ref{3.10}) using the approach
taken by Duval and Mitschi \cite{DM,Mi3,Mi2} based on obtaining the topological generators of the
Galois group, namely the formal and analytical invariants of the equation. We focus only on the
equation (\ref{3.10}) because the other equation (\ref{3.5}) after the change of the independent
variable $x = z^5 / (5^4 \lambda^4)$ becomes
\begin{gather*}
\delta \left(\delta + \frac{2}{5} -1\right) \left(\delta + \frac{3}{5} -1\right) \left(\delta + \frac{4}{5} - 1\right) \xi_1 - x \xi_1 = 0,
\end{gather*}
which is of similar kind as (\ref{3.10}).

The following equation
\begin{gather}
\label{4.2}
\mathrm{D}_{q p} (y) = \left[(-1)^{q-p} x \prod_{j=1} ^p (\delta + \mu_j) - \prod_{j=1} ^q (\delta + \nu_j -1) \right] y = 0,
\end{gather}
is called generalized conf\/luent hypergeometric equation since it generalizes the classical conf\/luent Kummer equation.
Here $\delta = x d/d x$, $0 \leq p \leq q$, $\mu_j, \nu_j \in \mathbb{C}$, $\mu_i - \mu_j \notin \mathbb{Z}$. For this equation
the point $0$ is a regular singularity and $\infty$ is an irregular singularity, assuming $p < q$.
For such kind of equations the local Galois group~$G_0$ is a subgroup of~$G_{\infty}$, so the global Galois group
is $G=G_{\infty}$.
In what follows we need some notations:
\begin{enumerate}\itemsep=0pt
\item[1)] $ \underline{\alpha} = (\alpha_1, \ldots, \alpha_n) \in \mathbb{C}^n $.

\item[2)] $ \langle\underline{\alpha}\rangle_m = \prod\limits_{j=1} ^n \alpha_j (\alpha_j + 1) \cdots (\alpha_j + m -1) $.

\item[3)] For $a \in \mathbb{C}^p, b \in (\mathbb{C} \setminus \mathbb{Z}^-)^q$, let
\begin{gather*}
{}_p F_q (\underline{a}; \underline{b} \, | x) = \sum_{n\geq0} \frac{\langle\underline{a}\rangle_n}{\langle\underline{b}\rangle_n} \frac{x^n}{n!}
\end{gather*}
be the generalized hypergeometric series and
\begin{gather*}
G_{p \, q} ^{m \,  n} \left (x \, \Big| {\underline{a} \above 0pt \underline{b}} \right) = \frac{1}{2 \pi i} \int_{C}
\frac{\prod\limits_{j=1} ^m \Gamma(b_j -s) \prod\limits_{j=1} ^n \Gamma(1 -a_j + s)}{\prod\limits_{j=m+1} ^q \Gamma(1-b_j +s) \prod\limits_{j=n +1} ^q \Gamma(a_j - s)}
x^s d   s ,
\end{gather*}
be the Meijer $G$-function \cite{Meijer} and  $C$ is a~suitable path in the complex plane.
It is proven that ${}_p F_q$ can be expressed as a~$G$-function.

\item[4)] Let $\widetilde{\mathbb{C}}$ be the Riemann surface of the logarithm, $l_1, l_2 \in \mathbb{R}\colon l_1 < l_2$.
 By a sector we understand the following set
\begin{gather*}
\theta (l_1, l_2) := \big\{ x \in \widetilde{\mathbb{C}} \,|\, \arg x \in (l_1, l_2) \big\}.
\end{gather*}

\item[5)] $ \sigma = q - p$, $\zeta = e^{2 \pi i /\sigma}$, $\lambda = \frac{1}{2}(\sigma +1) + \sum\limits_{j=1} ^p \mu_j - \sum\limits_{j=1} ^q \nu_j $.
\end{enumerate}

If $(\nu_j - \mu_k) \notin \mathbb{Z} $ for $j=1, \ldots, q$ and $k=1, \ldots, p$,
there exists a basis of solutions to (\ref{4.2}) near $x=0$ given in terms of $G_{p \, q} ^{1 \,  p}$ or ${}_p F_{q-1}$.
Similarly, in a neighborhood of $x = \infty$ there exists a~fundamental system of solutions expressed in terms of
$G_{p \, q} ^{q \,  1}$ and $G_{p \, q} ^{q \,  0}$ (see~\cite{Luke,Meijer}).
We specialize them for our particular case.

Duval and Mitschi have calculated the formal invariants (formal
monodromy and exponential torus) and the analytic invariants
(Stokes matrices) for all families of the equations~(\ref{4.2})
assuming $p \geq 1$. Their calculations can be adapted to the case
$p=0$ in which we are. It is obvious that the Kloosterman equation
is nothing but  a~$D_{q 0}$ type equation. Note that these equations
are generically irreducible over $\mathbb{C} (x)$ (see  \cite[p.~297]{BBH}
and \cite[p.~41]{DM} for the proof).

In the rest of the section we carry out detailed
calculations in our particular case for the reader's convenience.

Let us rewrite $L_2$ from (\ref{3.10}) in the form
\begin{gather*}
L_2 = (\delta + \nu_1 - 1) (\delta + \nu_2 - 1)(\delta + \nu_3 - 1)(\delta + \nu_4 - 1) - x,
\end{gather*}
which is  of type $\mathrm{D}_{4 \, 0}$ with $\nu_1 = 1$, $\nu_2 = -2/5$, $\nu_3 = 1/5$, $\nu_4 = 2/5$.

A fundamental system of solutions near $x=0$ of $L_2 \xi_1 = 0$ is
\begin{gather*}
\left\{ x^{7/5}{}_0 F_3 \left(;\frac{8}{5},\frac{9}{5},\frac{12}{5}\Big| x\right), x^{3/5}{}_0 F_3 \left(;\frac{1}{5},\frac{4}{5},\frac{8}{5}\Big| x\right), {}_0 F_3 \left(;-\frac{2}{5},\frac{1}{5},\frac{2}{5}\Big| x\right),\right.\\
\left.\qquad
x^{4/5} {}_0 F_3 \left(;\frac{2}{5},\frac{6}{5},\frac{9}{5}\Big| x\right)\right\}.
\end{gather*}
Then the monodromy $M_0$ around $x=0$ becomes
\begin{gather*}
M_0 = \mathrm{diag} \big(e^{2 \pi i 2/5}, e^{2 \pi i 3/5}, 1, e^{2 \pi i 4/5}\big).
\end{gather*}
Since 0 is regular singular, the monodromy generates the local Galois group~$G_0$.

Recall that in our case we have $\sigma = 4$, $\zeta = i$, $\lambda = 5/2 - 6/5$.
We prefer using letters~$\zeta$ and~$\lambda$ instead of their particular values.

Let us turn to the description of the local Galois group $G_{\infty}$.
To def\/ine a fundamental system near $x = \infty$ we need one more function. Let $C$ be a path in the complex
plane, connecting $-i \infty$ and $i \infty$ and enclosing the points $n - \nu_j$, $j = 1, \ldots, 4$, $n \in \mathbb{N}$.
The following function
\begin{gather}
\label{4.9}
\mathbb{G}_0 (x) := G_{0 \, 4} ^{4 \,  0} \left (x \, \Big| {\underline{} \above 0pt \underline{\nu}} \right) = \frac{1}{2 \pi i} \int_{C}
\Gamma (1 -\underline{\nu} - s) x^s d s
\end{gather}
is a solution of $L_2 \xi_1 = 0$, holomorphic in $\theta (-2 \pi, 2 \pi)$.
The analytic continuation of $\mathbb{G}_0$ in the sector $\theta(-5\pi, 5\pi)$ admits the following asymptotic expansion at
$x \to \infty$ (see \cite[Propositions~1.2 and~1.3]{DM} or~\cite{Luke})
\begin{gather}
\label{4.91}
e^{-4 x^{1/4}} x^{\lambda/4} \Theta (x),
\end{gather}
where $\Theta$ is a formal series in $x^{-1/4}$.
It is straightforward that
\begin{gather}
\label{4.92}
\mathbb{G}_0 \big(x e^{-2 \pi i h}\big), \qquad h \in \mathbb{Z}
\end{gather}
are also solutions of the same equation.
In order to get a fundamental solution near $x = \infty$, one needs to pick four of them in our case. Next, we need a
particular version of Meijer's expansion theorem which expresses every $G$-function as a f\/inite sum of the functions
$\mathbb{G}_0 (x e^{-2 \pi i h})$.

\begin{proposition}[see \protect{\cite[Proposition~1.5]{DM}}] \label{prop1}
 Let $x \in \theta (3 \pi, 5 \pi)$. Then the following identity holds
\begin{gather}
\label{4.10}
\mathbb{G}_0 (x) = A_1 \mathbb{G}_0 \big(xe^{-2\pi i}\big) + A_2 \mathbb{G}_0 \big(xe^{-4\pi i}\big) + B_0 \mathbb{G}_0 \big(xe^{-8\pi i}\big) + B_1 \mathbb{G}_0 \big(xe^{-6\pi i}\big) ,
\end{gather}
where
\begin{gather}
\label{4.11}
A_h = - \frac{d^h}{d x^h}\big(\langle 1 - x e^{-2 \pi i \underline{\nu}} \rangle_1 \big)_{x=0} , \qquad
B_h = e^{2 \pi i \lambda} \frac{d^h}{d x^h} \big(\langle 1 - x e^{2 \pi i \underline{\nu}} \rangle_1 \big)_{x=0} .
\end{gather}
\end{proposition}

The formal solutions of $D_{q p}$ at $\infty$ are known \cite{BBH,Luke} and can be verif\/ied by computer in our
particular case. It is more convenient from now on to use a new variable $t = x ^{1/4}$.
Denote by~$\underline{L}_{2}$ the equation obtained after the change of variable. Then we have
\begin{alignat*}{3}
&\underline{\Theta}_{-1} (t)  = e^{-4 \zeta^{-1} t} t^{\lambda} \underline{\Theta} \big(\zeta^{-1} t\big), \qquad&&
\underline{\Theta}_{0} (t)    = e^{-4 \zeta^{0} t} t^{\lambda} \underline{\Theta} \big(\zeta^{0} t\big),&  \\
&\underline{\Theta}_{1} (t)  = e^{-4 \zeta^{1} t} t^{\lambda} \underline{\Theta} \big(\zeta^{1}t\big), \qquad &&
\underline{\Theta}_{2} (t)  = e^{-4 \zeta^{2} t} t^{\lambda} \underline{\Theta} \big(\zeta^{2} t\big), \nonumber &
\end{alignat*}
where $\underline{\Theta} (t) \in \mathbb{C} [[t^{-1}]]$. We denote the basis of formal solutions at $\infty$ of $\underline{L}_{2}$
by
\begin{gather*}
\underline{\Sigma} (t) = \big\{ \underline{\Theta}_{-1} (t), \underline{\Theta}_0 (t), \underline{\Theta}_1 (t), \underline{\Theta}_2 (t) \big\}.
\end{gather*}
In this basis the formal monodromy is $\underline{\Sigma} (\zeta t) = \underline{\Sigma} (t) \widehat{M}_{\infty}$:
\begin{gather*}
\widehat{M}_{\infty} = e^{2 \pi i \lambda/4}
\begin{pmatrix}
0 & 0 & 0 & 1 \\
1 & 0 & 0 & 0 \\
0 & 1 & 0 & 0 \\
0 & 0 & 1 & 0
\end{pmatrix} .
\end{gather*}
Since in our case
\begin{gather}
\label{4.31}
q_{-1} (t) = -4 \zeta^{-1} t, \qquad  q_{0} (t) = - 4 \zeta^0 t, \qquad q_{1} (t) = -4 \zeta^1 t, \qquad q_{2} (t) = - 4 \zeta^2 t,
\end{gather}
the exponential torus is $\mathcal{T}\cong (\mathbb{C}^*)^2$ and can be presented by
\begin{gather*}
\mathcal{T} = \big\{ \mathrm{diag} \big(t_1, t_2, t_1 ^{-1}, t_2 ^{-1}\big) \big\}, \qquad t_1, t_2 \in \mathbb{C}^* .
\end{gather*}

The Stokes rays can easily calculated from (\ref{4.31}) to be
\begin{gather}
\label{4.33}
\arg t = n \frac{\pi}{4} , \qquad n = 0, \ldots, 7.
\end{gather}
Similarly the singular rays $d_s$, i.e., the rays bisecting the sectors
$\mathrm{Re} (q_i (t) - q_j (t)) < 0$ turn out to be the same as~(\ref{4.33}).

Let us def\/ine the sectors
\begin{gather*}
\theta _n = \theta \left(-\frac{\pi}{2} + \frac{n-1}{4} \pi, \frac{\pi}{2} + \frac{n}{4} \pi \right), \qquad n = 0, \ldots, 7 .
\end{gather*}

The following proposition is proven by Ramis~\cite{Ra2} for the general conf\/luent hypergeometric equation
$D_{q p}$. We reformulate it for our particular case.

\begin{proposition}
\label{prop2}
For every sector $\theta_n$, $n \in [0, 1, 2, \ldots, 7]$, there exists a unique basis of solu\-tions~$\Sigma_n (t)$ of
$\underline{L}_{2}$ in~$\theta_n$ with asymptotic expansion~$\underline{\Sigma} (t)$   at~$\infty$.
\end{proposition}

This solution corresponds to ``summation'' of~$\underline{\Sigma} (t)$ along a direction in the sector
$\theta \left(\frac{n-1}{4} \pi, \frac{n}{4} \pi \right)$. As we will see in the sequel we won't need
summation because there exist fundamental systems of actual solutions in~$\theta_n$.

In these notations the Stokes matrix corresponding to the singular ray $\frac{n}{4} \pi, \,  n \in [0, 1, \ldots, 7]$
is def\/ined via
\begin{gather}
\label{4.35}
\Sigma_n (t) = \Sigma_{n+1} (t) S_n, \qquad t \in \theta_n \cap \theta_{n+1} .
\end{gather}

\begin{proposition}
\label{prop3}
Suppose $n$ and $n'$ belong to $[0, 1, \ldots, 7]$ and $n' - n = 2$. Then
\begin{gather*}
\Sigma_{n'} (\zeta t) = \Sigma_n (t) \widehat{M}_{\infty} .
\end{gather*}
\end{proposition}

\begin{proof}
 If $t \in \theta_n$, then $\zeta t \in \theta_{n'}$ and
$\Sigma_{n'} (\zeta t) \widehat{M}_{\infty} ^{-1}$ is a basis of solutions to~$\underline{L}_2$ in~$\theta_n$ which admits the asymptotic expansion $\underline{\Sigma}_n (t)$. The uniqueness
from Proposition~\ref{prop2} gives the desired result.
\end{proof}

\begin{proposition}\label{prop4}
Let $n \in [0, 1, \ldots, 7]$ and $n = 2 m  + r$. Then
\begin{gather*}
S_n = \widehat{M}_{\infty} ^{-m} S_r \widehat{M}_{\infty} ^{m} .
\end{gather*}
\end{proposition}

\begin{proof}
From  the relation~(\ref{4.35}) after changing variables we obtain
$\Sigma_n (\zeta t) = \Sigma_{n+1} (\zeta t) S_n$. Proposition~\ref{prop3} gives that
\begin{gather*}
\Sigma_{n-2} (t) \widehat{M}_{\infty} = \Sigma_{n + 1 -2} (t) \widehat{M}_{\infty} S_n .
\end{gather*}
This procedure repeated $m$ times yields
\begin{gather*}
\Sigma_{r} (t) \widehat{M}_{\infty} ^m = \Sigma_{r + 1} (t) \widehat{M}_{\infty}^m S_n.
\end{gather*}
But by def\/inition we have $\Sigma_r (t) = \Sigma_{r+1} (t) S_r$ from where  the result is immediate.
\end{proof}

This proposition reduces the calculation of the Stokes matrices to $S_0$ and $S_1 : = S_{\pi/4}$ only.

\begin{proposition}
\label{prop5}
The function $V (t) = \mathbb{G}_0 (t^4)$ is asymptotic to $\underline{\Theta}_0$ in $\theta (-\frac{3\pi}{4}, \frac{5\pi}{4})$
when $t \to \infty$.
\end{proposition}

This is a reformulation of (\ref{4.9}) and (\ref{4.91}) in terms of the variable $t$.

\begin{proposition}\label{prop6}
If $t \in \theta(\frac{3\pi}{4}, \frac{5\pi}{4})$, the following identity holds
\begin{gather}
\label{4.36}
V (t) - e^{2 \pi i \lambda} V \big(t e^{-2 \pi i}\big) = A_1 V \big(t \zeta^{-1}\big) + A_2 V \big(t \zeta^{-2}\big)
+ B_1 V \big(t e^{-2 \pi i} \zeta\big) .
\end{gather}
\end{proposition}

This is a version of the formula (\ref{4.10}) in terms of the variable $t$
($B_0 = e^{2 \pi i \lambda}, \zeta = e^{2 \pi i / 4}$).

Next, we f\/ind fundamental systems of actual solutions near $\infty$ only in
$\theta_0$, $\theta_1$, $\theta_2$ since we need only $S_0$ and $S_1$.

\begin{proposition}[see \protect{\cite[Proposition 4.8]{DM}}]\label{prop7}
Let $n = 0, 1, 2$. The following sets of
solutions form fundamental systems of actual solutions in $\theta_n$:
\begin{gather*}
\Sigma_n (t) := \{Y_{n, j} (t), \, j \in \{-1, 0, 1, 2\} \},
\end{gather*}
where
\begin{gather}
Y_{n, -1} (t) = \zeta^{\lambda} V \big(t \zeta^{-1}\big), \qquad  n = 0, 1, 2 ,\nonumber \\
Y_{n, 0} (t)  =   V (t), \qquad   n = 0, 1, 2 , \nonumber\\
Y_{n, 1} (t)  =  \zeta^{-\lambda} V (t \zeta), \qquad n = 0, 1,\nonumber
\\
\label{4.40}
Y_{0, 2} (t) =
\begin{cases}
\zeta^{-2 \lambda} \big[ V \big(t \zeta^2\big) - A_1 V (t \zeta) \big], \quad & t \in \theta\big({-}\frac{3\pi}{4}, \frac{\pi}{4}\big) , \\
\zeta^{-2 \lambda} \big[ e^{2 \pi i \lambda} V \big(t e^{-2 \pi i} \zeta^2\big) + A_2 V (t) + B_1 V \big(t\zeta^{-1}\big) \big], \quad & t \in \theta\big({-}\frac{\pi}{4}, \frac{\pi}{2}\big),
\end{cases} \\
\label{4.41}
Y_{1, 2} (t)  =
\begin{cases}
\zeta^{-2 \lambda} \big[ V \big(t \zeta^2\big) - A_1 V (t \zeta) - A_2 V (t) \big], \quad &  t \in \theta\big({-}\frac{\pi}{2}, \frac{\pi}{4}\big) , \\
\zeta^{-2 \lambda} \big[ e^{2 \pi i \lambda} V \big(t e^{-2 \pi i} \zeta^2\big) + B_1 V \big(t\zeta^{-1}\big) \big], \quad & t \in \theta\big({-}\frac{\pi}{4}, \frac{3\pi}{4}\big) ,
\end{cases} \\
\label{4.42}
Y_{2, 1} (t)  =
\begin{cases}
\zeta^{- \lambda} \big[ V (t \zeta) - A_1 V (t) \big], \quad & t \in \theta\big({-}\frac{\pi}{4}, \frac{3\pi}{4}\big) , \\
\zeta^{- \lambda} \big[ e^{2 \pi i \lambda} V \big(t e^{-2 \pi i} \zeta\big) + A_2 V \big(t \zeta^{-1}\big) + B_1 V \big(t\zeta^{-2}\big) \big], \quad & t \in \theta\big(\frac{\pi}{4}, \pi\big),
\end{cases} \\
\label{4.43}
Y_{2, 2} (t)  =
\begin{cases}
\zeta^{-2 \lambda} \big[ V \big(t \zeta^2\big) - A_1 V (t \zeta) - A_2 V (t) - B_1 V \big(t\zeta^{-1}\big) \big], \quad & t \in \theta\big({-}\frac{\pi}{4}, \frac{\pi}{4}\big) , \\
\zeta^{-2 \lambda}  e^{2 \pi i \lambda} V \big(t e^{-2 \pi i} \zeta^2\big), \quad & t \in \theta\big({-}\frac{\pi}{4}, \pi\big) .
\end{cases}
\end{gather}
\end{proposition}

\begin{proof}
Rewriting the formula (\ref{4.92}) in terms of $t$ gives us solutions of $\underline{L}_{2}$
\begin{gather*}
V \big(t \zeta^{-h}\big), \qquad h \in \mathbb{Z}.
\end{gather*}
Using Proposition~\ref{prop6} we combine some of them in order to obtain proper asymptotic in~$\theta_n$.
 It remains to verify the validity of the
formulas (\ref{4.40})--(\ref{4.43}) in the intersection of their def\/inition intervals.
We check only (\ref{4.40}) since the rest are treated in the same way. So, we need to verify that
\begin{gather*}
V \big(t \zeta^2\big) - A_1 V (t \zeta) =  e^{2 \pi i \lambda} V \big(t e^{-2 \pi i} \zeta^2\big) + A_2 V (t) + B_1 V \big(t\zeta^{-1}\big)
\end{gather*}
is valid in $\theta(-\frac{\pi}{4}, \frac{\pi}{4})$. Using (\ref{4.36}) from Proposition \ref{prop6} ($\zeta^4 = e^{2 \pi i}$)
and making the change $t \to t\zeta^2$ gives the needed result.
\end{proof}

Denote by $E_{i j}$ the square matrix with elements 1 at $i$,\,$j$ place and zeroes elsewhere.
\begin{proposition}[see \protect{\cite[Theorem~5.2]{DM}}]\label{prop8}
The Stokes matrices $S_0$ and $S_1$ are given by the following formulas
\begin{gather*}
S_0  =  \mathbb{I} + \zeta^{-2\lambda} A_2 E_{2 4} ,  \qquad
S_1  =  \mathbb{I} + \zeta^{-\lambda} A_1 E_{2 3} + e^{-i \pi \lambda} \zeta^{-\lambda} B_1 E_{1 4} .
\end{gather*}
\end{proposition}

\begin{proof}
By def\/inition $\Sigma_n (t) = \Sigma_{n+1} (t) S_n$, $t \in \theta_n \cap \theta_{n+1}$. For
$t \in \theta_0 \cap \theta_{1} = \theta (-\frac{\pi}{2}, \frac{\pi}{2})$ from Proposition~\ref{prop7} we get
\begin{alignat*}{3}
& Y_{0, -1} (t)   = Y_{1, -1} (t), \qquad && Y_{0, 0} (t)   = Y_{1, 0} (t), & \\
& Y_{0, 1} (t)    = Y_{1, 1} (t), \qquad  && Y_{0, 2} (t)   = Y_{1, 2} (t) + \zeta^{-2 \lambda} A_2 Y_{1, 0} (t) .&
\end{alignat*}
Then
\begin{gather*}
\{Y_{0,-1} (t), Y_{0,0} (t), Y_{0,1} (t), Y_{0,2} (t)\} = \{Y_{1,-1} (t), Y_{1,0} (t), Y_{1,1} (t), Y_{1,2} (t)\} S_0
\end{gather*}
gives $S_0$. Similarly, for $t \in \theta_1 \cap \theta_{2} = \theta (-\frac{\pi}{4}, \frac{3 \pi}{4})$
we have
\begin{alignat*}{3}
& Y_{1,-1} (t) = Y_{2, -1} (t), \qquad && Y_{1, 0} (t)  = Y_{2, 0} (t), & \\
& Y_{1, 1} (t) = Y_{2, 1} (t)+\zeta^{-\lambda} A_1 Y_{2, 0},   \qquad && Y_{1, 2} (t)  = Y_{2, 2} (t) +
\zeta^{-2 \lambda} B_1 \zeta^{-\lambda} Y_{2, -1} (t) .&
\end{alignat*}
Then
\begin{gather*}
\{Y_{1,-1} (t), Y_{1,0} (t), Y_{1,1} (t), Y_{1,2} (t)\} = \{Y_{2,-1} (t), Y_{2,0} (t), Y_{2,1} (t), Y_{2,2} (t)\} S_1
\end{gather*}
gives $S_1$ since $\zeta^{-2\lambda} = e^{-\pi i \lambda}$.
\end{proof}

From the above proposition  we know the Stokes matrices $S_0$ and $S_{1}$.
In our case easy calculations give that
\begin{gather*}
S_0 = \mathbb{I} + a E_{2 4}, \qquad S_1 = \mathbb{I} + b E_{2 3} + c E_{1 4},
\end{gather*}
where $a = 2 i$, $c = i \zeta^{-\lambda} e^{2 \pi i /5}$, $b = i c$.

Then  Proposition \ref{prop4} gives the other Stokes matrices obtained from $S_0$, $S_1$:
\begin{gather*}
S_2 = \mathbb{I} + a E_{1 3}, \qquad S_4 = \mathbb{I} + a E_{4 2}, \qquad S_6 = \mathbb{I} + a E_{3 1}, \\
S_3 = \mathbb{I} + c E_{1 2} + i c E_{4 3}, \qquad S_5 = \mathbb{I} + c E_{4 1} + i c E_{3 2}, \qquad S_7 = \mathbb{I} + c E_{3 4} + i c E_{2 1}.
\end{gather*}

Now consider the (connected) subgroup topologically generated by the Stokes matrices and the exponential torus
$G_s = \{S_j, \mathcal{T} \}$ which is normal in the Galois group~$G_{\infty}$~\cite{Mi2}.
Hence, by Theorem~\ref{th5}, $G_{\infty}$ is topologically generated by $G_s$ and $\widehat{M}_{\infty}$. Let
 $\mathcal{G}_s$  be the Lie algebra of $G_s$ ($\mathcal{G}_s \subset  \mathrm{sl} (4, \mathbb{C}))$.

To compute $G_{\infty}$ we f\/irst determine the Lie algebra $\mathcal{G}_s$, then the corresponding connected subgroup
$G_s$ of $G_{\infty}$ and after that we describe the action of $\widehat{M}_{\infty}$ on $G_s$ (see \cite{Mi2}).

We will show that $\mathcal{G}_s \cong  \mathrm{sp} (4, \mathbb{C})$.
Denote $s_j \in \mathcal{G}_s$ such that $S_j = \exp  s_j$.
We have that $[s_2, s_6] = a^2 d_1$, $d_1 = E_{1 1} - E_{3 3}$ and  $[s_0, s_4] = a^2 d_2$, $d_2 = E_{2 2} - E_{4 4}$.
Then the Lie algebra $\mathcal{G}_s$ admits the following basis
\begin{gather*}
\mathcal{B} : = \{ s_0, s_2, s_4, s_6, d_1, d_2, s_1, s_3, s_5, s_7 \}.
\end{gather*}
Hence, $\mathcal{G}_s$ consists of all matrices $V$ such that $V^T J_1 + J_1 V = 0$, where
$J_1$ is the following skew-symmetric matrix
\begin{gather*}
J_1 = \begin{pmatrix}
0 & 0 & i \beta & 0\\
0 & 0 & 0     &  \beta \\
-i \beta & 0 & 0 & 0 \\
0 & - \beta & 0 & 0
\end{pmatrix}, \qquad \beta^4 = -1.
\end{gather*}
Therefore, we get that $G_s  \cong \mathrm{Sp} (4, \mathbb{C})$ (and $G^0 = \mathrm{Sp} (4, \mathbb{C})$).

Furthermore,  $(\widehat{M}_{\infty} ^k )^T J_1 \widehat{M}_{\infty} ^k \neq J_1$, $k = 1, 2 , 3 , 4$, i.e.,
$\widehat{M}_{\infty} ^k \notin G_s$, but $\widehat{M}_{\infty} ^5$ already belongs to~$G_s$.
Hence, the formal monodromy generates a f\/inite group $G_M$ isomorphic to $\mathbb{Z}/ 5 \mathbb{Z}$ acting nontrivially on~$G_s$.
This gives that~$G_{\infty}$ (and therefore~$G$) is isomorphic to a non-trivial semidirect product
$\mathrm{Sp} (4, \mathbb{C}) \rtimes \mathbb{Z}/ 5 \mathbb{Z}$.

\section[Non-integrability of the second and third members of the  $\mathrm{P}_{\rm II}$-hierarchy]{Non-integrability of the second and third members\\ of the  $\boldsymbol{\mathrm{P}_{\rm II}}$-hierarchy}\label{section5}

We have proved that some particular fourth-order Painlev\'e
equation is non-integrable in the Liouville sense for the set of
parameters $\gamma/\lambda = 3 k$ and  $\gamma/\lambda = 3 k - 1$,
$k \in \mathbb{Z}$. It turns out that the normal variational
equations along certain rational solutions are well-known
generalized hypergeometric equations whose dif\/ferential Galois
groups can be found. Since generically these groups are large, the
non-integrability comes from the Morales-Ruiz--Ramis theorem.

We  brief\/ly mention  an interesting relation concerning the
linear equations that have appeared in this paper. Let $X$ be a
smooth complex projective Fano variety. One can def\/ine quantum
dif\/ferential equations on $X$ (see, e.g.,~\cite{MvdP,Guest} and the
references there for details). When the quantum
equation is a linear ordinary dif\/ferential equation Cruz Morales and
 van der Put \cite{MvdP} conf\/irm Dubrovin's conjecture that the
Gram matrix of $X$ coincides with the Stokes matrix of the quantum
dif\/ferential equation (up to certain equivalence). It appears that
for $X=\mathbb{P}^{n-1}$ the quantum dif\/ferential operator is the Airy
type operator $\delta ^n - z$, and for the weighted projective
spaces $\mathbb{P} (w_0, \ldots, w_n)$ the quantum dif\/ferential
operator is of Kloosterman type or~$D_{q 0}$ for certain~$q$. The
classical Stokes matrices are then computed for these operators
using ``multisummation''  and the ``monodromy identity'' (see
\cite{MvdP} for details).

It is interesting to note that generalized hypergeometric functions and generalized conf\/luent
hypergeometric equations are also related  with other Painlev\'e equations. For instance,
the classical dilogarithm
\begin{gather*}
\mathrm{Li}_2 (z) = - \int_0 ^z \frac{\ln (1-s)}{s} d s ,
\end{gather*}
whose nontrivial monodromy plays an essential role in proving the non-integrability of some
Painlev\'e VI equations studied by Horozov and Stoyanova \cite[p.~626]{HSt} is related to the
generalized hypergeometric function as
\begin{gather*}
\mathrm{Li}_2 (z) = z \, {}_3F_2 (1,1,1;2,2 \,|\, z).
\end{gather*}
The polylogarithms $\mathrm{Li}_k$ have similar representations.

Let us turn our attention to other higher-order Painlev\'e equations
which admit a Hamiltonian formulation. Consider the $\mathrm{P}_{\rm II}$-hierarchy
which is given by (see~\cite{MazMo} and the references there)
\begin{gather}
\label{7.1}
\mathrm{P}^{(n)}_{\rm II}\colon \, \left(\frac{d}{d z} + 2 w \right)\!\mathcal{L}_n \big[w' -w^2\big] +
\sum_{l=1} ^{n-1} \beta_l \left(\frac{d}{d z} + 2 w \right)\!\mathcal{L}_l \big[w' -w^2\big] = z w + \alpha_n, \quad n\geq 1,\!\!\!
\end{gather}
where $\mathcal{L}_n$ is the operator def\/ined by the recursion relation (the Lenard relation)
\begin{gather}
\label{7.2}
\frac{d}{d z} \mathcal{L}_{n+1} = \left[\frac{d^3}{d z^3} + 4\big(w' - w^2\big) \frac{d}{d z} + 2 \big(w'-w^2\big)_z \right] \mathcal{L}_n, \qquad
\mathcal{L}_0 \big[w'-w^2\big] = \frac{1}{2}
\end{gather}
and $\beta_l$ and $\alpha_n$ are arbitrary complex parameters.
(Denoting for short $u:= u (z):= w' - w^2$, one gets consecutively $\mathcal{L}_1 [u] = u$, $\mathcal{L}_2 [u] = u'' + 3 u^2$,
  $\mathcal{L}_3 [u] = u^{(4)} + 10 u u'' + 5 (u')^2 + 10 u^3$
and so on).
A particular member of~(\ref{7.1}) is a nonlinear ODE of order~$2 n$, $n \geq 1$.
Some authors consider all~$\beta_l$ to be trivial. The f\/irst three members of the $\mathrm{P}_{\rm II}$-hierarchy are
\begin{alignat*}{3}
&   \mathrm{P}^{(1)}_{\rm II} \colon \ &&  w'' -2 w^3 = z w + \alpha_1, & \\
&   \mathrm{P}^{(2)}_{\rm II}\colon \ &&  w^{(4)} -10w\big(w w'' + w'^2\big) + 6w^5 + \beta_1 \big(w'' -2 w^3\big) = z w + \alpha_2, & \\
&  \mathrm{P}^{(3)}_{\rm II}\colon \ &&  w^{(6)}-14w^{(4)}w^2 -56w^{(3)}w' w + 70w''\big(w^4-w'^2\big) + 140w^3 w'^2 -42w (w'')^2  & \\
&  && {} -20w^7 + \beta_1 \big[w'' -2 w^3\big]  + \beta_2 \big[w^{(4)} -10w\big(w w'' + w'^2\big) + 6w^5\big] = z w + \alpha_3 . &
\end{alignat*}
The equation $\mathrm{P}^{(2)}_{\rm II}$ appears in~\cite[p.~58]{Cos} as F-XVII.

We are interested in the integrability of the Hamiltonian systems corresponding to these equations. The Hamiltonian
for  $\mathrm{P}^{(1)}_{\rm II}$ was known long ago from~\cite{Oka} (and also for the
other classical Painlev\'e equations).
The Hamiltonian structure for
the $\mathrm{P}_{\rm II}$-hierarchy was found by Mazzocco and Mo~\cite{MazMo}. We study the Liouville integrability
of the Hamiltonian systems corresponding to the f\/irst three members of the $\mathrm{P}_{\rm II}$-hierarchy, which are
``manageable''.

Consider f\/irst the Hamiltonian for  $\mathrm{P}^{(1)}_{\rm II}$,  namely
\begin{gather}
\label{7.6}
H^{(1)}= 4p^2 + \frac{1}{4}q + \frac{1}{4} pq^2 + 2 pz - \frac{1}{2} q \alpha_1,
\end{gather}
where $q = 4 w$, $p = \frac{1}{2}(w' - w^2 -\frac{z}{2})$. Extending~(\ref{7.6}) in a natural way to a two degrees of
freedom autonomous Hamiltonian system $\hat{H}_1 = H^{(1)} + F$, one  f\/inds $('=d/ds)$ the corresponding equations
\begin{gather}
\label{7.7}
 q'   =   8 p + \frac{1}{4} q^2 + 2 z, \qquad z' = 1, \qquad
 p'   =   -\frac{1}{4}-\frac{1}{2} p q + \frac{1}{2} \alpha_1, \qquad  F' = -2 p.
\end{gather}
The system (\ref{7.7}) admits the following particular solution when $\alpha_1 = 0$
\begin{gather}
\label{7.8}
q = 0, \qquad p = -\frac{1}{4} s, \qquad z = s, \qquad F = \frac{s^2}{4} + F_0.
\end{gather}
The NVE along (\ref{7.8}) is
\begin{gather*}
\xi_1 '' = z \xi_1 .
\end{gather*}
This is the Airy equation whose Galois group is $G = {\rm{SL}} (2, \mathbb{C}) \cong {\rm{Sp}} (2, \mathbb{C})$.
The Hamiltonian system (\ref{7.7}) is therefore non-integrable with rational f\/irst integrals, but we know that
from~\cite{MR3}.

Next we consider the Hamiltonian for $\mathrm{P}^{(2)}_{\rm II}$:
\begin{gather*}
H^{(2)} = \frac{q_2}{16} + 2z p_2 -16p_1 ^2 p_2 + 16 p_2 ^2 + \frac{q_1 q_2 p_2}{8} + \frac{p_1 p_2 q_2 ^2}{16}
+\frac{\alpha_2(p_1q_2-q_1)}{8} + \beta_1 (8p_1 - \beta_1)p_2,
\end{gather*}
where $q_j$, $p_j$, $j=1,2$ are expressible via $w$ and its derivatives. Extending as usual to a three
degrees of freedom autonomous Hamiltonian system $\hat{H}_2 = H^{(2)} + F$ we get
\begin{gather}
q_1 '  =  -32 p_1 p_2 + \frac{1}{16} p_2 q_2^2 + \frac{1}{8} q_2 \alpha_2 + 8 \beta_1 p_2, \nonumber \\
q_2 '  =  2 z -16 p_1 ^2 + 32 p_2 + \frac{1}{8} q_1 q_2 + \frac{1}{16} p_1 q_2 ^2 + \beta_1 (8 p_1 - \beta_1) , \nonumber \\
p_1 '  =  - \frac{1}{8} p_2 q_2 + \frac{1}{8} \alpha_2, \nonumber\\
p_2 '  =  -\frac{1}{16} - \frac{1}{8}p_2 q_1 -\frac{1}{8} p_1 p_2 q_2 - \frac{1}{8} \alpha_2 p_1, \nonumber \\
z'     =  1, \qquad F' = -2 p_2 . \label{7.11}
\end{gather}
The system (\ref{7.11}) admits the following particular solution when $\alpha_2 = 0$:
\begin{gather}
\label{7.12}
q_1 = q_2 = 0, \qquad  p_1 = \frac{\beta_1}{4}, \qquad p_2 = - \frac{s}{16}, \qquad z = s, \qquad F = \frac{s^2}{16} + F_0 .
\end{gather}
The NVE along the solution (\ref{7.12}) reduced to a single linear equation is
\begin{gather*}
\xi_1 ^{(4)} - \frac{5}{s} \xi_1 ^{(3)} + \left( \frac{12}{s^2} - \frac{\beta_1 s}{16} \right) \xi_1 '' +
\left( \frac{\beta_1}{16} - \frac{12}{s^3} \right) \xi_1 ' - \frac{s^3}{256} \xi_1 = 0.
\end{gather*}
Here we take the case $\beta_1 = 0$ which is simpler. After introducing the new independent variable
$z = s^7 /(2^8 7^4)$ the above equation becomes
\begin{gather}
\label{7.14}
\delta \left(\delta + \frac{2}{7} -1\right) \left(\delta + \frac{3}{7} -1 \right) \left(\delta + \frac{5}{7} -1 \right) \xi_1 - z \xi_1 = 0,
\end{gather}
which is an equation of type $D_{4 \, 0} \, \xi_1 = 0$ with $\nu_1 = 1$, $\nu_2 = 2/7$, $\nu_3 = 3/7$, $\nu_4 = 5/7$. In a similar way
as in Section~\ref{section4} (or referring to~\cite{Katz}) one  obtains that the identity component of the Galois group of (\ref{7.14}) is
$G^0 = {\rm{Sp}} (4, \mathbb{C})$, which is not commutative.
Hence, the Hamiltonian system
corresponding to the higher-order Painlev\'e equation~$\mathrm{P}^{(2)}_{\rm II}$ is not integrable in the Liouville sense.

Finally, let us write the Hamiltonian for $\mathrm{P}^{(3)}_{\rm II}$
\begin{gather*}
H^{(3)}   =   64 p_1 ^4 - 192 p_1^2 p_2 + 128 p_1 p_3 + \frac{1}{64}p_3 q_3 ^2 - \frac{1}{64} p_1 q_2 ^2+ 64 p_2 ^2
          -   \frac{1}{32} q_1 q_2 + 2 z p_1 + \frac{q_3}{64} \\
\hphantom{H^{(3)}   = }{} - \frac{1}{32} \alpha_3 q_3 + 8 \beta_1 (p_1^2 - p_2)
          +   \beta_2 (4 p_1 ^2 \beta_2 - 4 p_2 \beta_2 - 32 p_1 ^3 + 64 p_1 p_2 - 2 p_1 \beta_1).
\end{gather*}
Extending as usual to a four degrees of freedom autonomous Hamiltonian system $\hat{H}_3 = H^{(3)} + F$, we obtain
\begin{gather}
q_1 '   =   256 p_1^3 - 384 p_1 p_2 + 128 p_3 - \frac{q_2^2}{64} + 2 z + 16 p_1 \beta_1 + 8 p_1 \beta_2^2 - 96 \beta_2 p_1^2 + 64 p_2 \beta_2 - 2 \beta_1 \beta_2, \nonumber \\
q_2 '   =   -192 p_1^2 +128 p_2 - 8 \beta_1 - 4 \beta_2^2 + 64p_1 \beta_2, \nonumber \\
q_3 '   =   128 p_1 + \frac{1}{64} q_3 ^2 , \nonumber \\
p_1 '   =   \frac{1}{32} q_2, \nonumber\\
p_2 '   =   \frac{1}{32} p_1 q_2 + \frac{1}{32} q_1 , \nonumber \\
p_3 '   =   -\frac{1}{32} p_3 q_3 - \frac{1}{64} + \frac{1}{32} \alpha_3 , \nonumber \\
z '   =   1, \qquad F' = - 2 p_1. \label{7.16}
\end{gather}
Here we consider only the case $\beta_1 = \beta_2 = 0$. When $\alpha_3 = 0$ the system (\ref{7.16}) admits the following
particular solution
\begin{gather}
\label{7.17}
q_1 = q_2 = q_3 = 0, \qquad p_1 = p_2 = 0, \qquad p_3 = - \frac{s}{64}, \qquad z = s, \qquad F = F_0 = \operatorname{const}.
\end{gather}
The NVE along the solution (\ref{7.17}), reduced to a single linear equation becomes
\begin{gather*}
\xi_1 ^{(6)} - \frac{4}{s} \xi_1 ^{(5)} + \frac{12}{s^2} \xi_1 ^{(4)} - \frac{24}{s^3} \xi_1 ^{(3)} + \frac{24}{s^4} \xi_1 '' -s \xi_1 = 0.
\end{gather*}
After changing the independent variable by $x = s^7/ 7^6 $ we obtain
\begin{gather*}
\delta \left(\delta + \frac{1}{7} -1 \right)\left(\delta + \frac{2}{7} -1 \right)\left(\delta + \frac{3}{7} -1\right)
\left(\delta + \frac{4}{7} -1\right)\left(\delta + \frac{6}{7} -1\right)\xi_1 - x \xi_1 = 0,
\end{gather*}
which is an equation of type $D_{6 \, 0}   \xi_1 = 0$ with $\nu_1 = 1$, $\nu_2 = 1/7$, $\nu_3 = 2/7$, $\nu_4 = 3/7$, $\nu_5 = 4/7$, $\nu_6 = 6/7$.
We proceed in a similar way as in Section~\ref{section4}. Here
\begin{gather}
\label{7.20}
\sigma = q-p = 6, \qquad \zeta = e^{2 \pi i/6}, \qquad \lambda = \frac{1}{2}- \frac{2}{7}.
\end{gather}
The local monodromy at~0 is clear. It is more convenient to use a new variable $t = x^{1/6}$. The basis of
the formal solutions is straightforward (see for instance \cite{BBH,DM}) and the formal monodromy~is
\begin{gather*}
\widehat{M}_{\infty} = e^{2 \pi i \lambda/6}
\begin{pmatrix}
0 & 0 & 0 & 0 & 0 & 1 \\
1 & 0 & 0 & 0 & 0 & 0 \\
0 & 1 & 0 & 0 & 0 & 0 \\
0 & 0 & 1 & 0 & 0 & 0  \\
0 & 0 & 0 & 1 & 0 & 0 \\
0 & 0 & 0 & 0 & 1 & 0
\end{pmatrix} .
\end{gather*}
The exponential torus is again $\mathcal{T}\cong (\mathbb{C}^*)^2$
\begin{gather*}
\mathcal{T} = \big\{ \mathrm{diag} \big(t_1, t_2, t_1 ^{-1} t_2, t_1 ^{-1}, t_2 ^{-1}, t_2 ^{-1} t_1 \big) \big\}, \quad t_1, t_2 \in \mathbb{C}^* .
\end{gather*}
The Stokes rays and the singular rays are
\begin{gather*}
\arg t = n \frac{\pi}{6} , \qquad n = 0, \ldots, 11.
\end{gather*}
We def\/ine the sectors
$
\theta _n = \theta \left(-\frac{\pi}{2} + \frac{n-1}{6} \pi, \frac{\pi}{2} + \frac{n}{6} \pi \right)$, $n = 0, \ldots, 11$.
Again we need only $S_0$ and $S_1 := S_{\pi/6}$ in order to obtain all Stokes matrices.
The Stokes matrices~$S_0$ and~$S_1$ are given by (see \cite[Theorem~5.2]{DM})
\begin{gather*}
S_0 = \mathbb{I} + a E_{4 5} + b E_{3 6} + c E_{2 1}, \qquad S_1 = \mathbb{I} + d E_{3 5} + f E_{2 6},
\end{gather*}
where $a = \zeta^{-\lambda} A_1$, $b = \zeta^{-3\lambda} A_3$, $c =  \zeta^{\lambda} e^{-2 \pi i \lambda} B_1$,
$d = \zeta^{-2\lambda} A_2$, $f = \zeta^{-\lambda} e^{- \pi i \lambda} B_2 $.
Using~(\ref{7.20}) and~(\ref{4.11}) we get
\begin{gather*}
f = i \zeta^{-\lambda} e^{2 \pi i 2/7}, \qquad d = f^2, \qquad c = \frac{i}{f}, \qquad b = i, \qquad a = i f, \qquad f^3 = -1.
\end{gather*}
The  matrices $S_0$ and $S_1$ together with Proposition \ref{prop4} for $n = 0, \ldots, 11$ give the rest of the Stokes
matrices
\begin{gather*}
 S_2  = \mathbb{I} + i f E_{3 4} + i E_{2 5} + \frac{i}{f}  E_{1 6}, \qquad S_3 = \mathbb{I} + f^2 E_{2 4} + f  E_{1 5}, \\
S_4 = \mathbb{I} + i f E_{2 3} + i E_{1 4} + \frac{i}{f}  E_{6 5}, \qquad
 S_5  = \mathbb{I} + f^2 E_{1 3} + f E_{6 4}, \\
  S_6 = \mathbb{I} + i f E_{1 2} + i E_{6 3} + \frac{i}{f}  E_{5 4},
\qquad S_7 = \mathbb{I} + f^2 E_{6 2} + f E_{5 3},     \\
 S_8  = \mathbb{I} + i f E_{6 1} + i E_{5 2} + \frac{i}{f}  E_{4 3}, \qquad S_9 = \mathbb{I} + f^2 E_{5 1} + f E_{4 2},  \\
 S_{10}  = \mathbb{I} + i f E_{5 6} + i E_{4 1} + \frac{i}{f}  E_{3 2}, \qquad S_{11} = \mathbb{I} + f^2 E_{4 6} + f E_{3 1}.  \nonumber
\end{gather*}

Now consider again the (connected) subgroup topologically generated by the Stokes matrices and the exponential torus
$G_s = \{S_j, \mathcal{T} \}$.
 Let  $\mathcal{G}_s$  be the Lie algebra of $G_s$ ($\mathcal{G}_s \subset  \mathrm{sl} (6, \mathbb{C}))$.

\begin{proposition}\label{prop9}
$\mathcal{G}_s \cong  \mathrm{sp} (6, \mathbb{C}) $.
\end{proposition}

\begin{proof}
Denote again $s_j \in \mathcal{G}_s$ such that $S_j = \exp  s_j$, $j = 0, \ldots, 11$
and $\tau_1 = E_{1 1} - E_{3 3} - E_{4 4} +  E_{6 6}$ and $\tau_2 = E_{2 2} + E_{3 3} - E_{5 5} -  E_{6 6}$
which belong to Lie~$\mathcal{T}$. Direct calculations yield
\begin{gather*}
 [s_0, s_3] = 2 i E_{2 5}, \qquad [s_0, s_9] = -2 i E_{4 1}, \qquad [s_6, s_9] = 2 i E_{5 2},\\
{}  [s_2, s_{11}] = -2 i E_{3 6}, \qquad [s_2, s_5] = 2 i E_{1 4}, \qquad [s_4, s_7] = 2 i E_{6 3},
\end{gather*}
and hence, $ E_{1 4}, E_{4 1}, E_{2 5}, E_{5 2}, E_{3 6}, E_{6 3} \in \mathcal{G}_s$.

Additionally we have that the elements
\begin{gather*}
B_1 := E_{1 1} - E_{4 4} = [s_8, s_2] - \tau_2, \qquad B_2 := E_{2 2} - E_{5 5} = [s_9, s_3] - B_1, \\
B_3 := E_{3 3} - E_{6 6} = B_1 - \tau_1,
\end{gather*}
and
\begin{gather*}
B_4 :=\frac{1}{f} E_{1 2} - E_{5 4} = \frac{i}{f^2} \left( [s_6, \tau_2] - 2 i E_{6 3} \right), \\
B_5 :=\frac{1}{f} E_{2 3} - E_{6 5} = - i f^2 \left( [\tau_1, s_4] - 2 i E_{1 4} \right),
\\
B_6 := f E_{2 1} - E_{4 5} = -i f^2 \left( [\tau_2, s_0] - 2 i E_{3 6} \right), \\
B_7 := f E_{3 2} - E_{5 6} = -i f^2 \left( [s_{10}, \tau_1] - 2 i E_{4 1} \right),
\\
B_8 :=\frac{1}{f^2} E_{1 6} + E_{3 4} = \frac{i}{f} \left( [\tau_2, s_2] - 2 i E_{2 5} \right), \\
B_9 :=\frac{1}{f^2} E_{6 1} + E_{4 3} = - i f \left( [s_8, \tau_2] - 2 i E_{5 2} \right)
\end{gather*}
also belong to $\mathcal{G}_s$.

Then the Lie algebra $\mathcal{G}_s$ admits the following basis :
\begin{gather*}
\mathcal{B} : = \{E_{1 4}, E_{4 1}, E_{2 5}, E_{5 2}, E_{3 6}, E_{6 3}, s_1, s_3, s_5, s_7, s_9, s_{11}, B_j, j=1, \ldots, 9 \}.
\end{gather*}
Hence, $\mathcal{G}_s$ consists of all matrices $V$ such that $V^T J_1 + J_1 V = 0$, where
$J_1$ is the following skew-symmetric matrix
\begin{gather*}
J_1 = \begin{pmatrix}
0 & 0 & 0 & f^2 & 0 & 0 \\
0 & 0 & 0 & 0   & f & 0 \\
0 & 0 & 0 & 0   & 0 & 1 \\
-f^2 & 0 & 0 & 0    & 0 & 0  \\
0 & -f & 0 & 0 & 0 & 0 \\
0 & 0 & -1 & 0 & 0 & 0
\end{pmatrix}
\end{gather*}
from where we get the desired result.
\end{proof}

Therefore, we get that $G_s  \cong \mathrm{Sp} (6, \mathbb{C})$ (and $G^0 = \mathrm{Sp} (6, \mathbb{C})$).

Furthermore,  $(\widehat{M}_{\infty} ^k )^T J_1 \widehat{M}_{\infty} ^k \neq J_1$, $k = 1, \ldots, 6$, i.e.,
$\widehat{M}_{\infty} ^k \notin G_s$, but $\widehat{M}_{\infty} ^7$ already belongs to~$G_s$.
Hence, the formal monodromy generates a f\/inite group $G_M$ isomorphic to $\mathbb{Z}/ 7 \mathbb{Z}$ acting nontrivially on~$G_s$.
In this way we obtain that $G_{\infty}$ (and therefore $G$) is isomorphic to a non-trivial semidirect product
$\mathrm{Sp} (6, \mathbb{C}) \rtimes \mathbb{Z}/ 7 \mathbb{Z}$.

Hence, the Hamiltonian system corresponding to the higher-order Painlev\'e equation $\mathrm{P}^{(3)}_{\rm II}$ is
not integrable in the Liouville sense.
Summarizing we have the following
\begin{theorem}\label{th7}
Suppose that
\begin{enumerate}\itemsep=0pt
\item[$(i)$] $\beta_1 = \alpha_2 = 0$. Then the Hamiltonian system corresponding to $\mathrm{P}^{(2)}_{\rm II}$
is not integrable by means of rational f\/irst integrals;

\item[$(ii)$] $\beta_1= \beta_2 = \alpha_3 = 0$. Then the Hamiltonian system corresponding to $\mathrm{P}^{(3)}_{\rm II}$
is not integrable by means of rational first integrals.
\end{enumerate}
\end{theorem}

The study of the other members of the $\mathrm{P}_{\rm II}$-hierarchy is technically involved.
However, we think that the NVE along certain nontrivial solutions reduced to  single
equations are of the type $D_{q \, 0}   \xi = 0$ with~$q$ even. Since the identity components
of their dif\/ferential Galois groups are  $\rm{Sp} (q, \mathbb{C})$,
the autonomous Hamiltonian systems corresponding to these equations are non-integrable.

The result of Theorem~\ref{th7} can be extended to the entire orbits of parameters using B\"acklund transformations
and other special solutions recently found by Sakka~\cite{Sak1,Sak2}. This issue will be addressed elsewhere.

\subsection*{Acknowledgements}

The authors are grateful to the referees for their constructive
criticism and suggestions.
 We also would like to thank Ivan Dimitrov for many useful discussions.
O.C.~acknowledges partial support by Grant 059/2014 with Sof\/ia
University.

\pdfbookmark[1]{References}{ref}
\LastPageEnding

\end{document}